\pdfoutput=1
\documentclass{aa}
\usepackage{graphicx}
\usepackage{natbib}
\usepackage{mydblfloatfix}

\usepackage{txfonts}
\usepackage{color}

\newcommand{\kms}{\,km\,s$^{-1}$}
\newcommand{\dg}{$^{\circ}$} 
\newcommand{\rs}{$\mathrm{R_{\odot}}$} 

\begin{document} 

\title{Tracking solar wind flows from rapidly varying viewpoints by 
the Wide-field Imager for Parker Solar Probe}
\titlerunning{Tracking solar wind flows with PSP WISPR}

\date{Received: Accepted: }

% \abstract{}{}{}{}{} 
% 5 {} token are mandatory
 
  \abstract
 {}
{Our goal is to develop methodologies to seamlessly track transient solar wind flows viewed by coronagraphs or heliospheric imagers from rapidly varying viewpoints.}
{We constructed maps of intensity versus time and elongation (J-maps)
from Parker Solar Probe (PSP) Wide-field Imager (WISPR) observations during the fourth encounter of PSP. From the J-map, we built an intensity on impact-radius-on-Thomson-surface map (R-map). Finally, we constructed a latitudinal intensity versus time map (Lat-map). Our methodology satisfactorily addresses the challenges associated with the construction of such maps from data taken from rapidly varying viewpoint observations.}
{Our WISPR J-map exhibits several tracks, corresponding to transient solar wind flows ranging from a coronal mass ejection (CME) down to streamer blobs. The latter occurrence rate is about 4-5 per day, which is similar to the occurrence rate in a J-map made from $\sim1$ AU data obtained with the Heliospheric Imager-1 (HI-1) on board the Solar Terrestrial Relations Observatory Ahead spacecraft (STEREO-A). STEREO-A was radially aligned with PSP during the study period. The WISPR J-map tracks correspond to angular speeds of $2.28 \pm 0.7$\dg/hour ($2.49 \pm 0.95$\dg/hour), for linear (quadratic) time-elongation fittings, and radial speeds of about 150-300 \kms. The analysis of the Lat-map reveals a bifurcating streamer, which implies that PSP was flying through a slightly folded streamer during perihelion.}
{We developed a framework to systematically capture and characterize transient solar wind flows from space platforms with rapidly varying vantage  points. The methodology can be applied to PSP WISPR observations as well as to upcoming observations from instruments on board the Solar Orbiter mission.}

   \keywords{Sun: solar wind -- Sun: corona}
\author{A. Nindos \inst{1}
\and S. Patsourakos \inst{1}
\and A. Vourlidas \inst{2}
\and P. C. Liewer\inst{3} \and  P. Penteado\inst{3} \and
J. R. Hall \inst{3}}
\institute{Physics Department, University of Ioannina, Ioannina GR-45110,
Greece\\
\email{anindos@uoi.gr}
\and
The Johns Hopkins University Applied Physics Laboratory, Laurel, MD 20723, USA
\and
Jet Propulsion Laboratory, California Institute of Technology, Pasadena, CA, 91109,
USA}
\authorrunning{A. Nindos et al.}
   \maketitle
%
%________________________________________________________________

\section{Introduction}

Solar wind, that is a flow of plasma from the solar corona which
plays a major role in shaping the heliosphere, was theorized in the late
1950s by  \citet[][]{parker1958}  and was one of the milestone
discoveries  during the dawn of the space age
\citep[][]{gringauz1960,neugebauer1962}.  A zeroth-order
classification of the solar wind is in terms of fast and slow solar
wind, with this distinction being eminent particularly during  solar
minimum conditions \citep[e.g.,][]{mccomas2000}.  The fast solar wind
is quasi-steady and mainly emanates from polar coronal holes, while
the slow solar wind is more variable, and mainly originates from
streamers.  Despite major progress \citep[e.g., see the reviews
  by][and references therein]{antiochos2012,cranmer17,vial2020},
there are still significant gaps in our understanding of the origins
and evolution of the  slow solar wind. It is unclear
whether the slow wind is predominately steady, quasi-steady, or intermittent and what the corresponding spatio-temporal scales associated with its release are. More often than not, the variability of the slow solar wind is interpreted in terms of its intermittent release during reconnection events between closed and open magnetic field lines
\citep[e.g.,][]{wang1998,antiochos2011,higginson2018}.

Remote-sensing observations of the corona and inner heliosphere were,
until very recently, the major means for probing the young solar wind. The standard instruments to perform such measurements are coronagraphs
and heliospheric imagers. Both imagers typically obtain observations over wide bandpasses in the visible. The observed emission is the result of Thomson scattering of  photospheric radiation from the free electrons in the corona (K-corona) and from interplanetary dust (F-corona). For the study of the corona and inner heliosphere in general, and for solar wind flows in particular, we need to isolate the K-corona. Since the corona is optically-thin in the visible spectral range, the K-corona intensity recorded by coronagraphs and heliospheric imagers includes contributions along the line-of-sight (LoS). However, given the
properties of Thomson scattering, the location of maximum scattering
efficiency lies on the so-called Thomson Surface
\citep[][]{vourlidas2006}, % howard2012,inhester2015}} 
which is a sphere with a diameter equal to the Sun-observer distance. For small elongations, the Thomson surface is equivalent to the traditional plane-of-sky concept \citep[][]{vourlidas2006}. 

Important results regarding the nature of the slow solar wind emerged
from the analysis of observations from the coronagraphs of the
Large Angle and Spectroscopic Coronagraphs
\citep[LASCO;][]{brueckner1995} on board the Solar and Heliospheric Observatory mission \citep[SOHO;][]{domingo1995} and more recently from the coronagraphs and heliospheric imagers of  the Sun-Earth Connection Coronal and Heliospheric Investigation 
\citep[SECCHI;][]{howard2008} instrument suite on-board the twin spacecraft of the Solar TErrestrial RElations Observatory mission
\citep[STEREO;][]{kaiser2008}. \citet{sheeley1997} and \citet{wang1998}
discovered a fragmented release of mass in streamers, in the form of 
blobs. The height-time plots of the blobs seem to follow standard
theoretical profiles of the slow solar wind.  Around four streamer
blobs per day are observed during solar minimum conditions
\citep[e.g.,][]{sheeley2009}. The blobs, and more generally, the
transient slow solar wind structures exhibit quasi-periodic behavior
with periods of $\approx 6-18$ hours
\citep{viall2010,viall2015,sanchez2017}.   The analysis of high
cadence, deep exposure coronagraph campaign SECCHI COR2 data by
\citet{deforest2018} has revealed an enhanced fine structure, intermittency,
and inhomogeneity in streamer flows.

The kinematics of streamer blobs, as well as of other transient flows in the solar wind, including  Coronal Mass Ejections (CMEs), are frequently studied via special time-distance maps, colloquially named ``J-maps'', introduced by \citet{sheeley1999}, mapping the intensity across angular elongation (that is, the angular distance from Sun center) sector versus time, for a narrow radial slice corresponding to a specific position angle \citep[e.g.,][]{davies2009,rouillard2011,lugaz2009,liu2010,sheeley2010}. J-maps allow easier  tracking of faint features than using plain images, and selected feature tracks (i.e., coherent collections of time-elongation pairs) can be inverted to yield  properties of the transient solar flows, including speed, acceleration, and direction. 

Until the launch of the Parker Solar Probe
\citep[PSP;][]{fox2016} mission on August 12 2018, imaging 
observations of the solar wind were available only from around 1 AU. PSP is the first  mission to fly into the
Sun's corona. Taking advantage of a sequence of Venus flybys,  the PSP
is diving progressively deeper into the Sun's corona from about 35
\rs\ down to 9.8 \rs\ for 24 planned perihelia during the nominal
mission.
\begin{figure*}[t]
\centering
\includegraphics[width=0.6\textwidth]{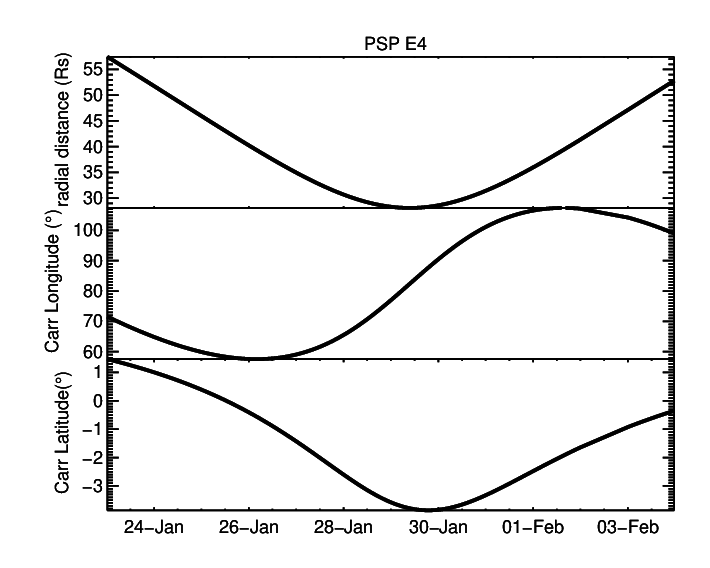}
\caption{PSP orbital information during the fourth perihelion (E4): Heliocentric radial distance (top panel), Carrington longitude (middle panel), and Carrington latitude (bottom panel).}
\label{orbit}
\end{figure*}
The Wide-Field Imager \citep[WISPR;][]{vourlidas2016} on board PSP,
is the first remote sensing instrument to view the corona  from
vantage points within the corona. Such coronal observations have two key advantages over those taken from 1 AU, namely, (i) less
``contaminating'' F-corona signal, and (ii) less integration along the LoS and more sensitivity to structures closer to the spacecraft
allowing to better probe the fine structure. 

The first WISPR results include the detection of a depletion of the
F-corona close to the Sun
\citep[][in this issue]{howard2019,stenborg2020} 
%\citep[][and Stenborg et al. 2020 in this issue]{howard2019},  
the detection of fine  structure in streamers
\citep[][]{poirier2020}, the kinematics of a pristine CME
\citep[][]{hess2020} including a J-map capturing its  evolution, a
study of the internal structure and force balance in that CME
\citep[][]{rouillard2020a}, and the kinematics and 3D morphology of a streamer blob \citep[][]{wood2020}.

However, the advantages of up-close WISRP observations come
with a significant cost: the rapidly varying vantage point of PSP, particularly during its perihelia,  give rise to a series of non-trivial challenges for the data analysis \citep[see][]{liewer2019}. For example, the rapidly changing LoS injects significant ambiguity in the interpretation of
running difference images, which are customarily used in the analysis of
data from the more ``stable'' 1~AU viewpoints. 
A detailed discussion about the challenges in constructing J-maps from a rapidly varying viewpoint and our strategy to mitigate them is given in Section \ref{jmapc}.

We hereby report on observations of transient solar wind flows during the fourth encounter of PSP (E4) in January, 2020. In Section~2, we present our observations and in Section~3 we outline the methodology for deriving various distance-time maps, including traditional J-maps, that can be used in the study of transient solar wind flows from observations obtained by rapidly varying viewpoints, with application to WISPR. In Section 4 we discuss the  resulting maps. Finally, Section 5 contains our conclusions and an outlook of  our study. 

\section{Observations}
\begin{figure*}[h]
\centering
\includegraphics[width=0.9\textwidth]{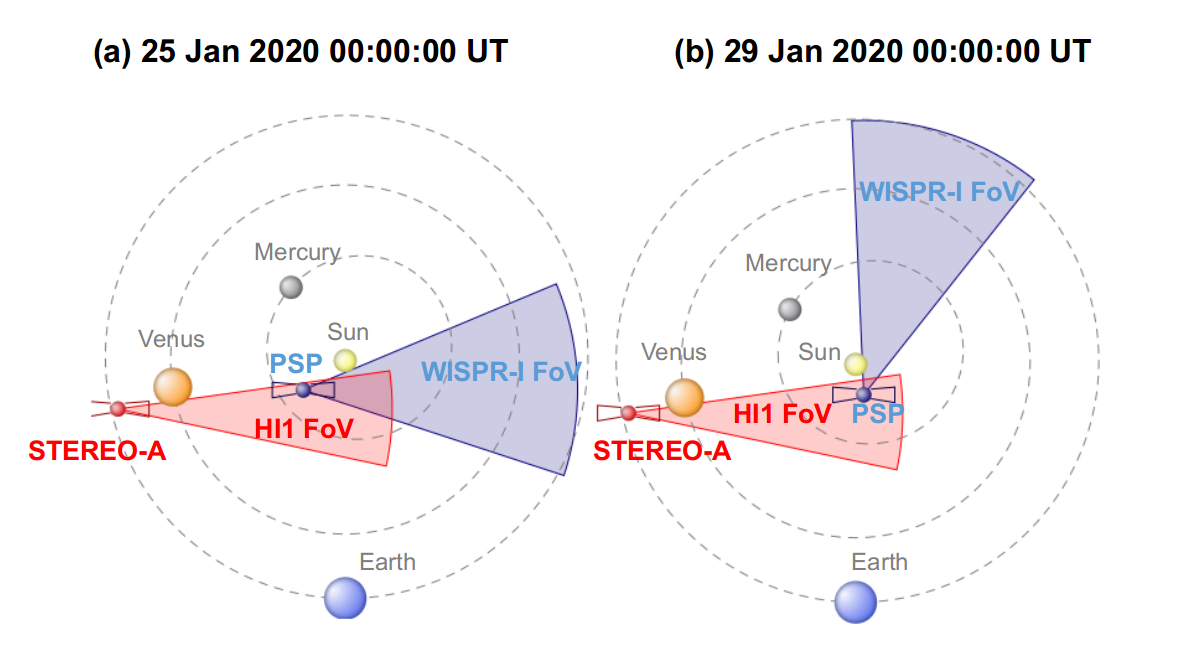}
\caption{Positions of STEREO-A (red filled circle) and PSP (blue filled circle) spacecraft together with fields of view of STEREO-A H1 (pink wedge)  and PSP WISRP-I (blue wedge) imagers on 25 January 00:00:00 UT (panel a) 
and 29 January 00:00:00 UT (panel b). This figure has been produced using the Propagation Tool described in \citet{rouillard2017}.}
\label{fov}
\end{figure*}
The observations were taken during the fourth encounter of PSP between 2020-01-23T00:44:26 UT and 2020-02-03T23:23 UT. Fig.
\ref{orbit} shows the  variation of the PSP radial distance, Carrington longitude, and latitude during E4. The PSP heliocentric distance varied from 55 \rs\ to the perihelion distance of 28 \rs\,  reached on 29 January, 2020 (Fig.
\ref{orbit}, top panel). Although the angular width of the WISPR field of view (FOV) remains constant, the change in radial distance changes dramatically the physical size of the coronal region recorded in the images  \citep{vourlidas2016,liewer2019}.

In the middle panel of Fig. \ref{orbit}, we note
 two periods of PSP quasi-corotation with the Sun (that
is, periods with constant PSP Carrington longitude)  in the inbound
($\approx$ 26-27 January, 2020) and outbound ($\approx$ 1-2 February, 2020) phase of E4. During these periods PSP is essentially ``locked'' with the same structures.  These two intervals of quasi-corotation encompass the period of PSP ``super-rotation''. During this period, PSP essentially flies into structures rooted at the Sun.  Finally, PSP stayed at or close the ecliptic plane during E4 (Fig.~\ref{orbit}, bottom panel). 

WISPR consists of a pair of telescopes (inner telescope, WISPR-I, and outer telescope, WISPR-O) mounted on the ram-side of PSP  \citep{vourlidas2016}. WISPR captures total brightness
images of the corona and the inner heliosphere in the spectral ranges of 490–740 nm (WISPR-I) and  475–725 nm (WISPR-O).  WISPR-I has a radial FOV spanning $\approx$ 13.5-51\dg\  in
elongation while its transverse FOV spans  58\dg. WISPR-I's FOV as
well as its positions at two instances during E4 are  presented in
Fig. \ref{fov}. For comparison, we show in the same figure  the
corresponding FOV and positions of the STEREO-A HI-1.  Interested readers are referred to the article by \citet{vourlidas2016} for more information about WISPR.

The nominal image cadence for WISRP-I during E4 was around 20
minutes. The  ``raw'' (L1) images were first
submitted to a number of  operations to convert instrument to physical  units (Mean Solar Brightness Units; MSB). These operations included corrections for geometric distortion, vignetting, and stray light followed by photometric
calibration leading to the generation of the so-called ``L2''
images. More information about WISPR calibration can be found in the WISP 
website (https://wispr.nrl.navy.mil/wisprdata). Next, the background F-corona emission was
removed from the L2 images, following a variation of the procedure outlined by \citet{stenborg2017}, which  essentially calculates a separate background image for each WISRP-I L2 image.  This approach is  absolutely necessary for images taken from rapidly varying
viewpoints  and led to  ``L3''
images. We note that the \citet{stenborg2017} procedure  besides
subtracting the F-corona, may also subtract some part of the
large-scale component of the K-corona as well.  Finally, the
background-subtracted L3 images were submitted  to spatial median
filtering in order to remove the influence of stars. Our further processing (see Section 3) is done on the L3 images.

\section{Construction of maps}

\subsection{J-maps}
\label{jmapc}
We first identify visually and discard several  ``bad'' images contaminated by bright streaks due to dust impacts onto the PSP spacecraft. Then, we apply the following steps.

\begin{figure*}
\centering
\includegraphics[width=0.99\textwidth]{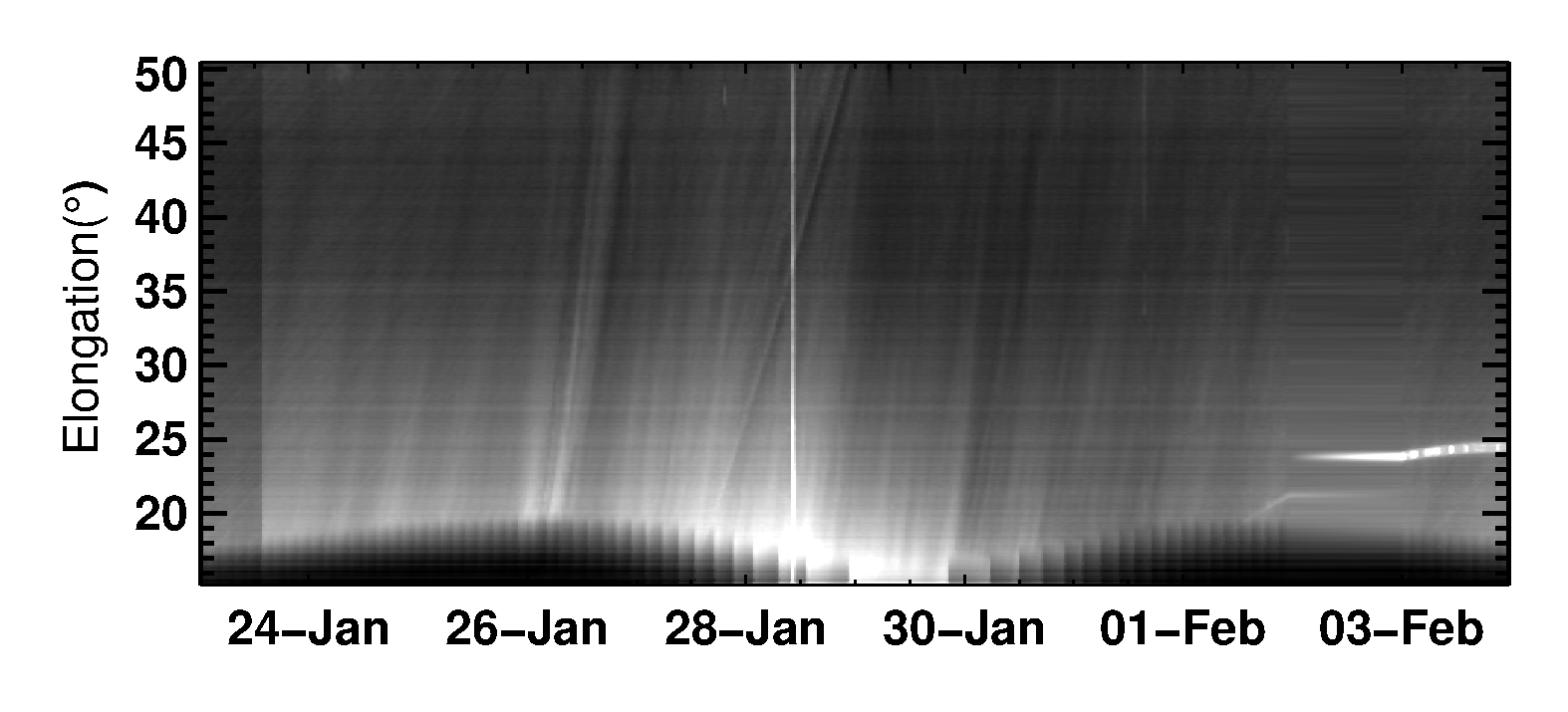}
\caption{WISPR-I J-map (intensity versus time and elongation) during PSP E4, after the application of steps 1-3 of Section \ref{jmapc}. The J-map was calculated using a conical sector with a full width of 1\dg; the vertex of the sector was at the PSP orbit plane and its axis of symmetry was parallel to that plane.}
\label{jmap2}
\end{figure*}

\begin{figure*}
\centering
\includegraphics[width=0.9\textwidth]{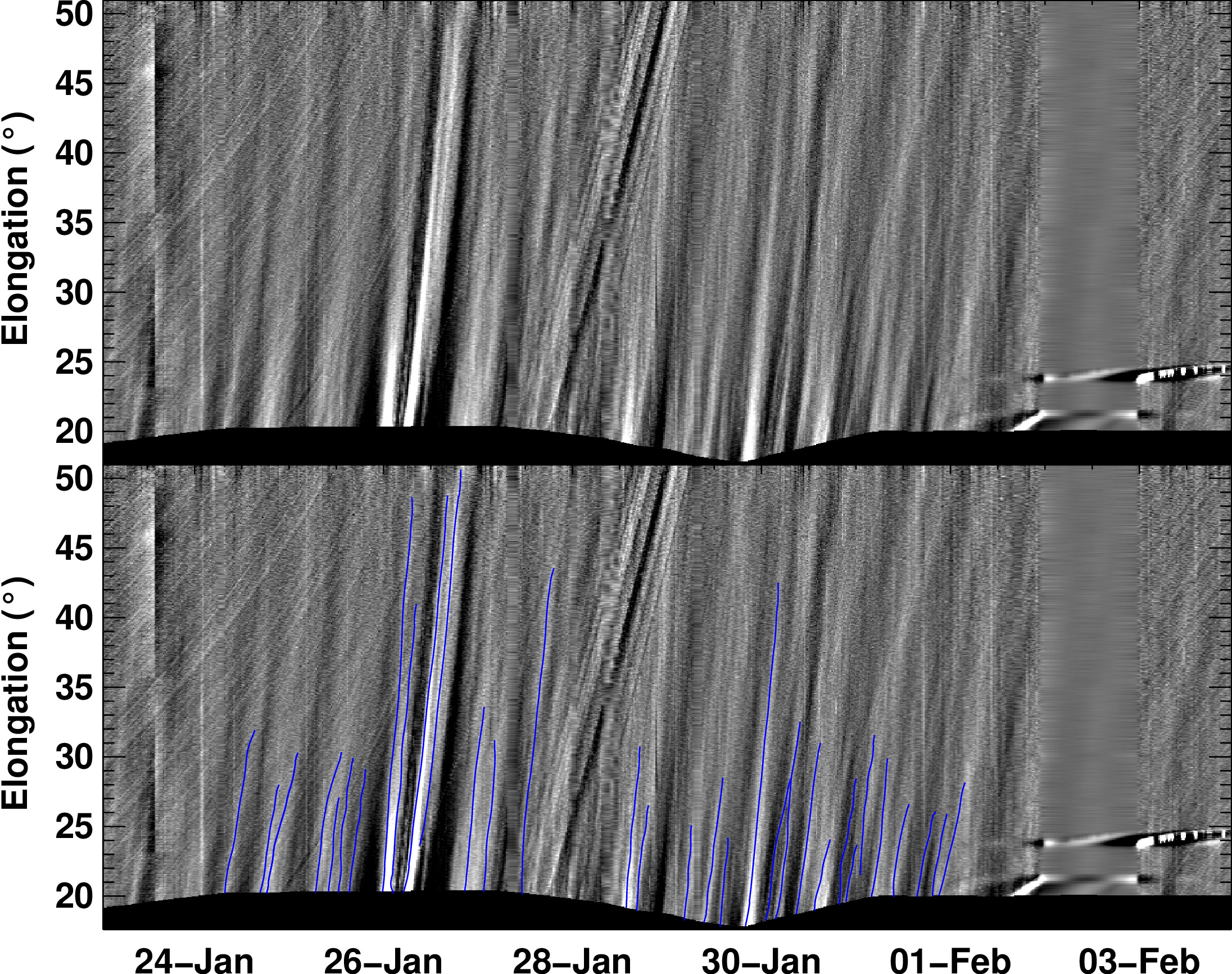}
\caption{Top panel: Same as Fig. \ref{jmap2}, but now after the application of all steps of our method. Bottom panel: same as top panel with the addition that several manually-traced tracks have been marked (blue curves).}
\label{jmap1}
\end{figure*}

\begin{figure*}[t]
\centering
\includegraphics[width=0.99\textwidth]{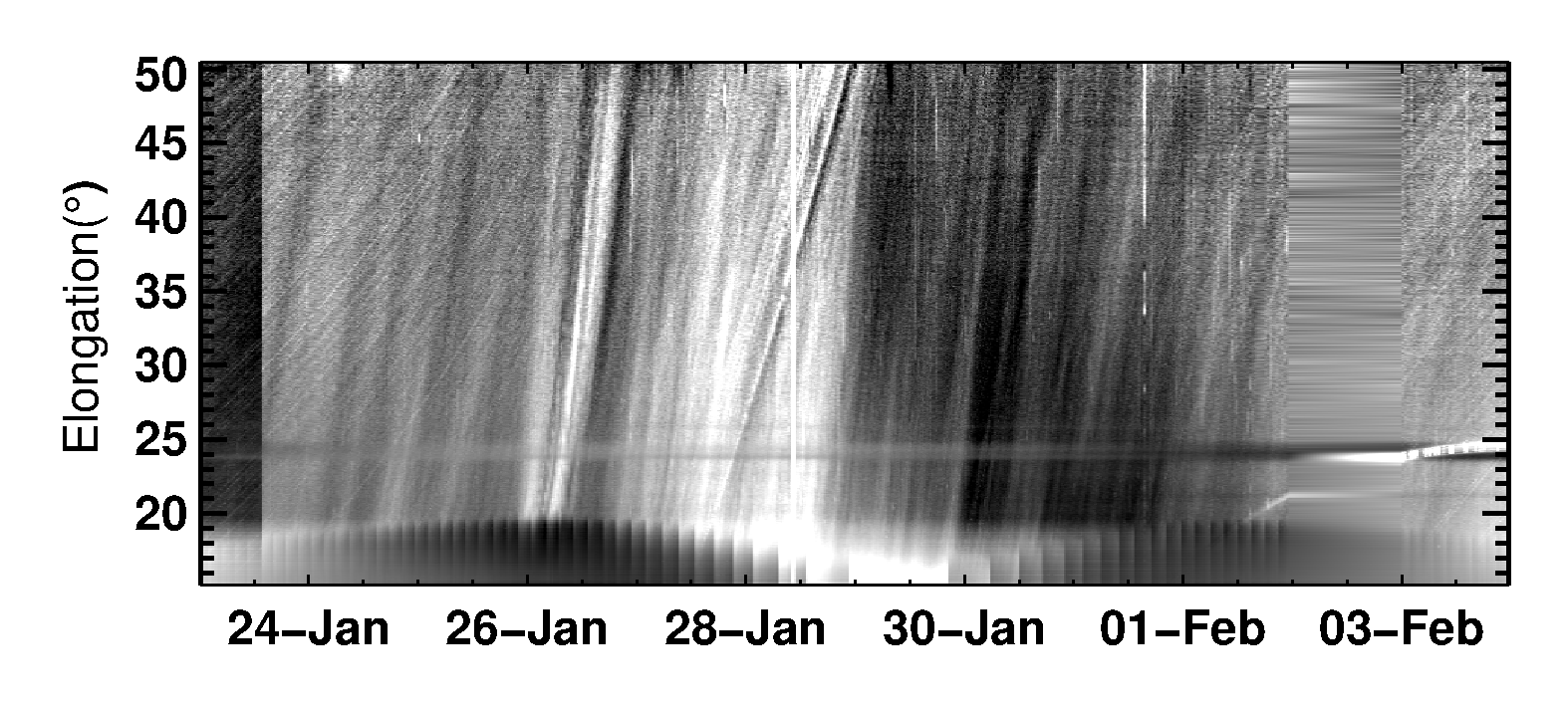}
\caption{WISPR-I J-map after replacing the last step of our standard methodology
with the subtraction of a second-order polynomial fit from each J-map light 
curve and then apply the Normalized Radial Graded Filter to the J-map rows.}
\label{jmap3}
\end{figure*}

The first step is the mapping of L3 images onto the PSP orbit frame.  
The WISPR-I L3 images represent the background-subtracted K-corona in detector coordinates. For a fixed (or slowly-varying) viewpoint, it is straightforward to map coronagraph or heliospheric-imager images to a heliocentric frame of reference or conversely to position angle. The latter type of projection is employed in the construction of J-maps from images obtained from SOHO or STEREO (see related references in the Introduction).  However, because of PSP's rapidly-changing elliptical orbit, features moving radially no longer remain at a fixed position angle. This is why J-maps can only be constructed from images projected on the so-called PSP orbit frame \citep[e.g.,][]{liewer2019,hess2020}.
The PSP orbit frame is
similar to the helio-projective Cartesian coordinate (HPC) frame \citep[][]{thompson2016}; like the HPC system, it is observer-centric but the two vectors that determine the PSP orbit plane are the Sun-spacecraft vector and the spacecraft velocity vector. The third axis is perpendicular to this plane. Thus, the 
columns of the projected image correspond to solar elongation while the rows correspond to constant latitude with respect to the PSP orbital plane.

In the second step we construct the J-maps.
To construct the ``standard'' J-maps discussed in the Introduction, we can take simply row averages around zero latitude and stack them in time. We note that this approach works only for features moving in the orbit plane. To increase the signal-to-noise ratio of the J-maps for easier tracking toward higher elongations, we use a conical sector with 1$^\circ$ angular width \citep[as in ][]{sheeley1999}. The use of a conical sector instead of a rectangular box increases the number of pixels with elongation, and therefore boosts the signal-to-noise at larger elongations. In addition, this approach is consistent with mass conservation along streamlines since the sector cross-section increases as the square of distance while the density falls off as the inverse square of distance. Comparisons of J-maps made with  sectors and rectangular boxes showed similar values for small
elongations, but a 30\% increase in the signal-to-noise
ratio for sector averaging at larger elongations. 
 
The third step is the mapping of J-maps into a uniform temporal grid.
To account for the nonuniform image cadence (due to either 
nonuniform image acquisition or to missing images), the J-maps from the previous steps are linearly interpolated onto a uniform time grid.

The fourth step involves the removal of orbital trend and contrast enhancement.
For each elongation, the variation of J-map intensity versus time (hereafter referred to as ``J-map light curve'') is
%The light curves in any given pixel are 
subject to a long-term orbital trend due to the varying PSP viewpoint. The effect can be understood as follows: since PSP is orbiting rapidly, a given pixel traces varying lines of sight through the coronal scene, resulting in rapidly varying
intensities. Smaller (larger) heliocentric distances correspond to shorter
(longer) line-of-sight impact parameters and hence lower (higher) resulting intensities. To remove this trend, we apply the Savitzky-Golay (SG) filter  \citep{Savitzky1964} to the J-map light curves. The SG filter is used widely in digital data processing and astrophysics \citep[e.g.,][]{bell1979,szkody1986,kozarev2017,barucci2020}. The filter essentially smooths the input data via polynomial least-square fitting, and therefore preserves the input data dynamic range, in contrast to boxcar filters which tend to reduce dynamic range because they average the data over sliding windows. Since the orbital trend we needed to remove was smooth with long timescale, we used a zero-order SG filter. 
The degree of the filter was chosen to emulate the variation of the PSP radial distance during E4 (see top panel of Fig. \ref{orbit}). Our experimentation showed that a fourth degree filter provided the best results.

The J-map resulting from steps 1-3 is shown in Fig. \ref{jmap2}. 
The top panel of Fig. \ref{jmap1} shows the J-map from the same data after the application of all steps (1-4) of our method. The black areas at the bottom part of the J-map are the result of blanking out those J-map pixels that were associated with areas outside the imaging ares of the WISPR detector (Fig. \ref{jmap2}). 
These pixels contain no scientific information and discarding them is essential for our method to work.

\begin{figure*}[b]
\centering
\includegraphics[width=0.99\textwidth]{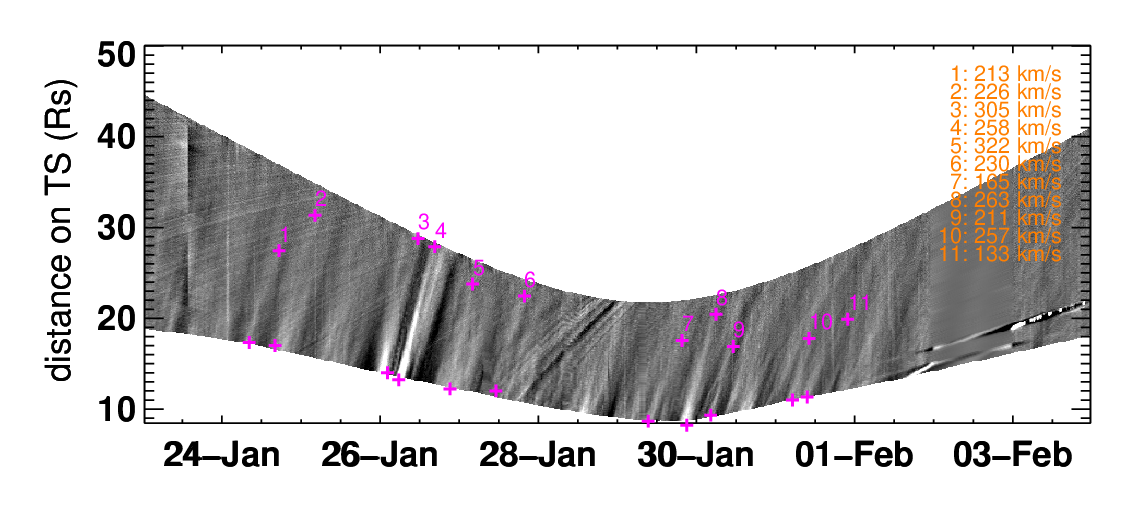}
\caption{WISPR-I R-map (intensity versus time and Thomson Surface radial distance expressed in \rs)  during PSP E4. The R-map was calculated from the J-map of Fig. \ref{jmap1} following the procedure outlined in Section \ref{rmapc}. The start and end points of several numbered tracks are marked with pink crosses and the speeds resulting from linear fittings of the associated time-distance pairs are displayed in the right upper part of the plot.}
\label{rmap}
\end{figure*}

\begin{figure*}[h]
\centering
\includegraphics[width=0.9\textwidth]{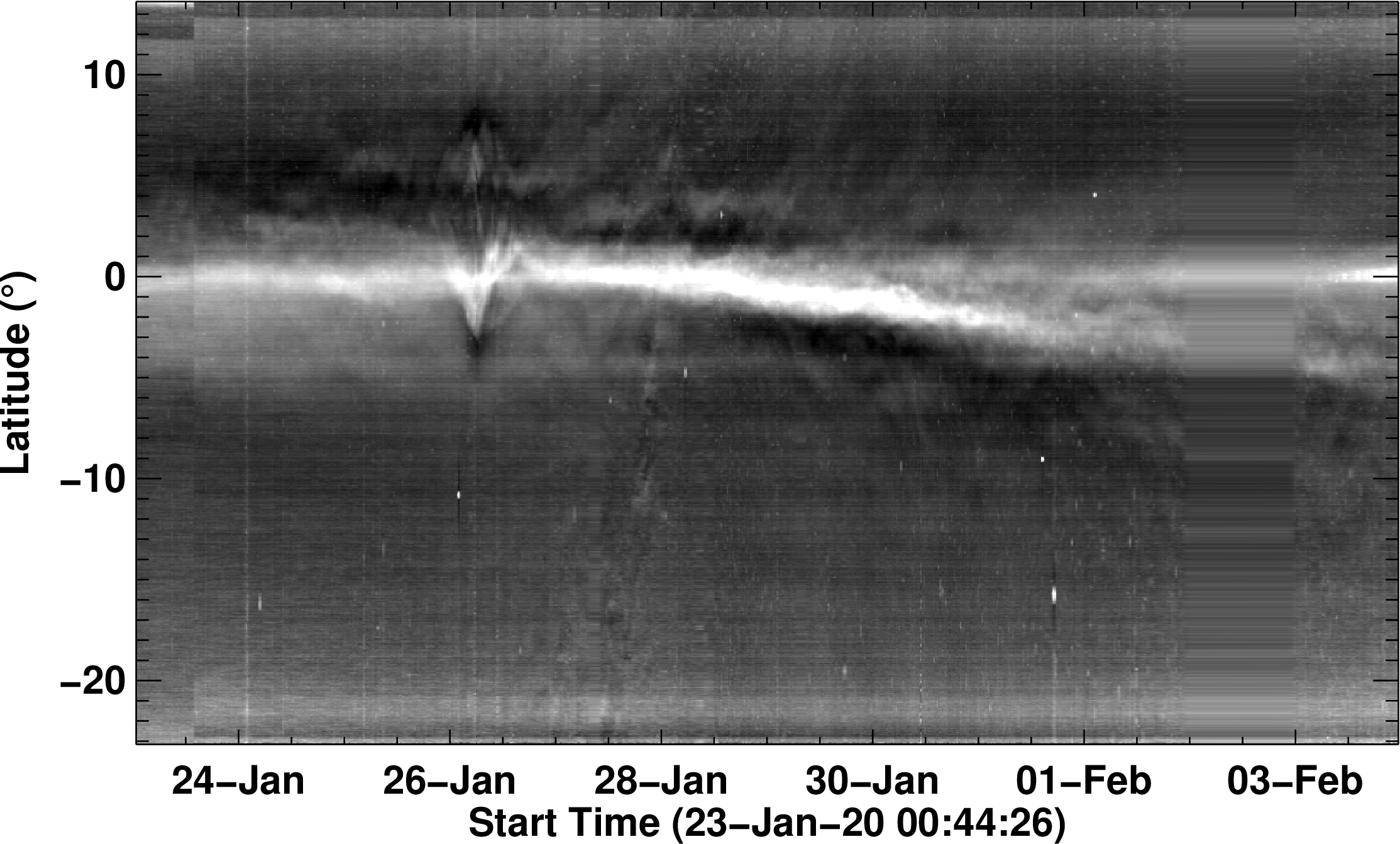}
\caption{WISPR-I Lat-map (intensity versus time and latitude) during PSP E4.
The Lat-map was calculated at an elongation of 25\dg.}
\label{latmap}
\end{figure*}

\begin{figure*}
\centering
\includegraphics[width=0.99\textwidth]{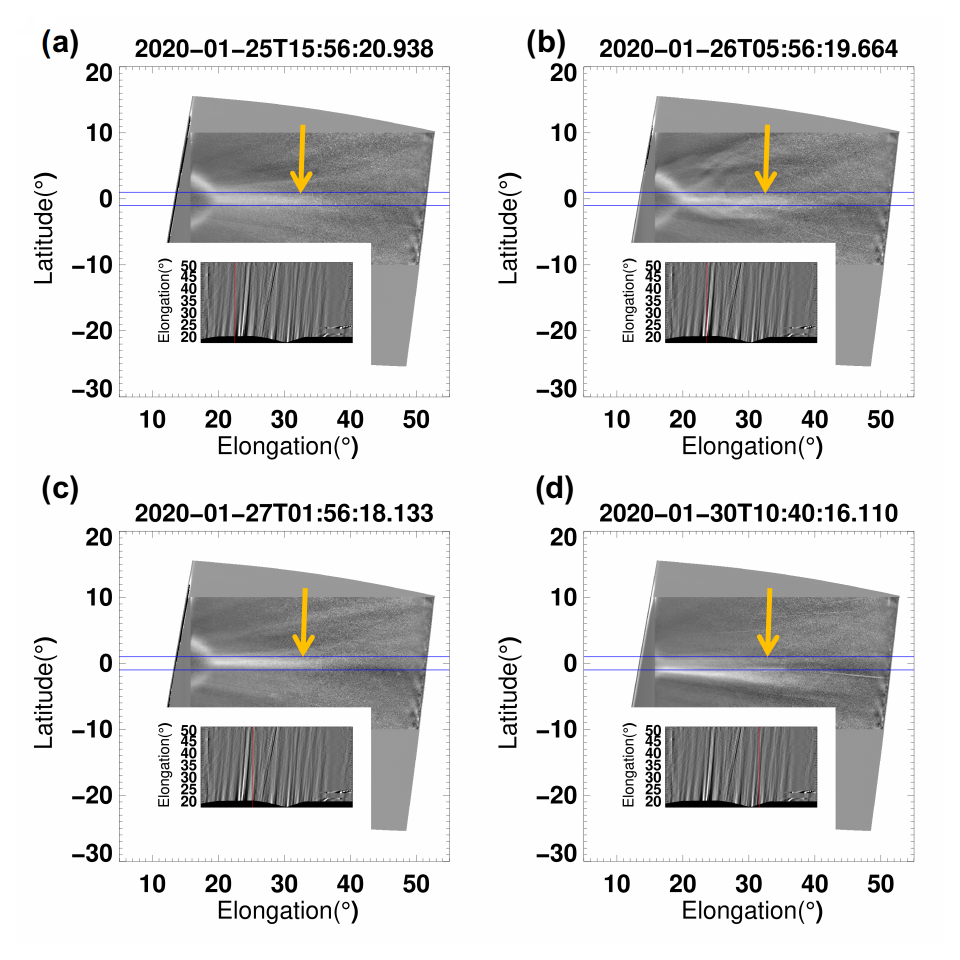}
\caption{WISPR-I images obtained during E4 after they have been projected 
onto the PSP orbit plane. The horizontal blue lines correspond to latitudes of $\pm 1$\dg\  from the PSP orbit plane, and are used to guide the eye into the region around zero latitude. The thick orange arrows mark the fronts of outgoing features. In each panel the insert shows the J-map of Figure~\ref{jmap1}; the vertical red line in each J-map insert marks the time of the corresponding WISPR image. See also the associated movie.}
\label{movjmap}
\end{figure*}

In the following, unless specified otherwise, we present results from the J-maps produced with the method discussed above, which constitutes our  ``base'' methodology. A spin-off of this methodology consists of replacing step 4 with the following procedure. First we subtract a second-order polynomial fit from each J-map light curve. Then, we apply the Normalizing Radial Graded Filter \citep[NGRF;][]{morgan2006} to the rows in J-maps.
From each row we first subtract the corresponding average value and then divide by the corresponding standard deviation. The NGRF essentially calculates variance-normalized residuals and hence is  insensitive to the intrinsic row intensity.
The resulting J-map is presented in Fig. 5.

A visual inspection of Figs. \ref{jmap1} and \ref{jmap3} indicates
that both approaches result in similar J-maps (see also Section 4.1) but our base methodology results in more uniform (that is, it shows less artifacts) and somewhat crisper maps. These arguments are confirmed by comparing the intensity root mean square (rms) of the two J-maps, as well as the width of their autocorrelation functions, respectively. The intensity rms of 
the J-map of Fig. 4 was 32\% smaller than the intensity rms of the J-map of
Fig. 5. Furthermore, the width of the autocorrelation function corresponding to the J-map of Fig. \ref{jmap1} was 21\% narrower than that of the NGRF-treated J-map of Fig. \ref{jmap3}.  
%This is confirmed by comparing the width of the autocorrelation functions of the two J-maps; the one corresponding to the J-map of Figure \ref{jmap1} was 21\% narrower than the NGRF-treated J-map of Figure \ref{jmap3}. 
 
Finally, we note that the application of the SG and the high-pass
filtering directly to the ``J-map space'' (that is, the J-maps themselves) instead of the original images, is less sensitive to noise because J-maps result from averaging in space. This is important for tracking features at large elongations in the J-maps.

\subsection{R-maps}
\label{rmapc}

Deriving radial distances from WISPR J-maps and images requires
geometrical models  \citep[see][]{liewer2019,nistico2020}. As a
zeroth-order approach, and  relying upon the properties of Thomson
surface, discussed in the Introduction, we introduce a new type of map, which we call ``R-map''. It is essentially a remapping of the angular distance to impact radius on the Thomson surface versus time. The impact radius corresponds to the ``d'' parameter in Fig.~1 of \citet{vourlidas2006}, and represents the heliocentric distance of the point on the Thomson surface. 
The following steps are applied to a J-map to built an R-map: (a) Convert elongations to impact radii on the Thomson surface using the corresponding PSP heliocentric distance and the geometry of Fig. 1 of  \citet{vourlidas2006}, and (b) Map the J-map onto a time-impact-radii-on-Thomson-surface grid. Since R-maps show intensity versus time in linear units, it is more straightforward to derive the kinematics of observed tracks from such maps. The R-map we produced is
presented in Fig. \ref{rmap}.

\subsection{Lat-maps}
\label{latmapc}

If, rather than slices in elongation versus time for a given latitude, we stack slices in latitude versus time at a given elongation, we obtain latitudinal J-maps, which we refer to as ``Lat-maps''. 
Such Lat-maps, introduced by \citet{sanchez2017}, track the
large-scale evolution of coronal and heliospheric structures at a
direction roughly perpendicular to the solar wind outflow. 
The motion of a feature on the Lat-map will be a combination of the spacecraft and feature's own motions \citep[][]{liewer2019}. Once the spacecraft motion is accounted for, these maps can, in principle, be used to determine feature motions in the same way as the elongation J-maps. As an example, we present a Lat-map at an elongation of 25\dg\ (Fig. \ref{latmap}).

\section{Discussion}

\subsection{J-maps}

The J-maps essentially display  coronal intensity versus time as a function of elongation along the PSP orbit plane. 
We can identify several tracks in the top panel of Fig.~\ref{jmap1}; some of them are marked in the bottom panel of the same Figure. The marked blue curves 
result from spline interpolation of pairs of manually-selected points along those tracks. The 
tracks  can also be identified, albeit with a lower contrast, on the ``raw'' J-map  of Fig. \ref{jmap2}, and also on the J-map of Fig. \ref{jmap3}. The close correspondence among the tracks 
identified in Figs.~\ref{jmap2}, \ref{jmap1} and
\ref{jmap3} demonstrates that our contrast enhancing procedure  does not  introduce any significant artifacts in the J-maps.

To validate the reliability of the J-maps, we compare J-maps features to structures in L3 WISPR movies. Such a movie accompanies Fig. \ref{movjmap}. In it we show WISPR-I images projected onto the PSP orbit frame and contrast-enhanced by applying the NGRF filter. The displayed images cover the full elongation range of WISRP-I and a sector of $\pm$ 10\dg\ from the PSP orbit plane. In each movie frame, the J-map of Fig. \ref{jmap1} appears as an insert containing a red vertical line that marks the time of the corresponding WISPR image. 
 
Apart from the Milky Way that enters the WISPR-I FOV on 27 January 2020, the most prominent drifting feature, in both the movie and the J-map, is a CME that enters the FOV on 26 January 2020  
\citep[an in-depth analysis of this CME is reported by][in this issue]{Liewer2020}. In addition to the CME, the movie shows numerous transient solar wind flows. They include several blobs as well fronts without a clear indication of a blob morphology. These morphologies seem consistent with a flux rope structure observed edge-on and face-on, respectively \citep[][]{sheeley2009}. Studies of streamer blobs observed by WISPR can be found in 
\citet[][]{howard2019} and in \citet[][]{wood2020}. The movie demonstrates the overall correspondence between transient flows and J-map tracks %(see Figure \ref{movjmap} for characteristic snapshots from the movie).
Characteristic snapshots from the movie are presented in Fig. \ref{movjmap}. Panels (a) and (c) correspond to times when outflows associated with the southern branch of the streamer were in progress while in panel (d) an outflow from the northern branch of the streamer was in progress. The image of panel (b) has been taken during the propagation of the CME. The J-map tracks that correspond to these outflows are crossed by the vertical red line that appears in the insert of each panel.

We now turn to a more detailed discussion of the J-map tracks in Fig. \ref{jmap1}.  First, most of the observed tracks 
exhibit a small curvature, reminiscent of the 
forward models of WISPR J-maps in \citet[][]{liewer2019}; their Fig. 9 features  
curved J-map tracks in the WISPR-I FOV
corresponding to two radially propagating small flux
ropes: one approaching PSP, and another approaching and 
intercepting PSP. Second, during E4 we can identify 4-5 tracks  per
day. Third, the average ``duration'' (that is, crossing time) of the longest tracks is around one day. Fourth, first and second order polynomial fits of the manually-traced tracks (bottom panel
of Fig. \ref{jmap1}), result in elongation rates of $2.28 \pm
0.7$\dg/hour and $2.49 \pm 0.95$\dg/hour, respectively. These values do not correspond directly to kinematic properties of the outflows. The similarity of the results from the linear and quadratic fittings indicates that the curvature of the tracks is small, in agreement with the visual inspection of Fig. 4. 
This implies that the outflows experience little acceleration.  

\begin{figure*}[h]
\centering
\includegraphics[width=0.8\textwidth]{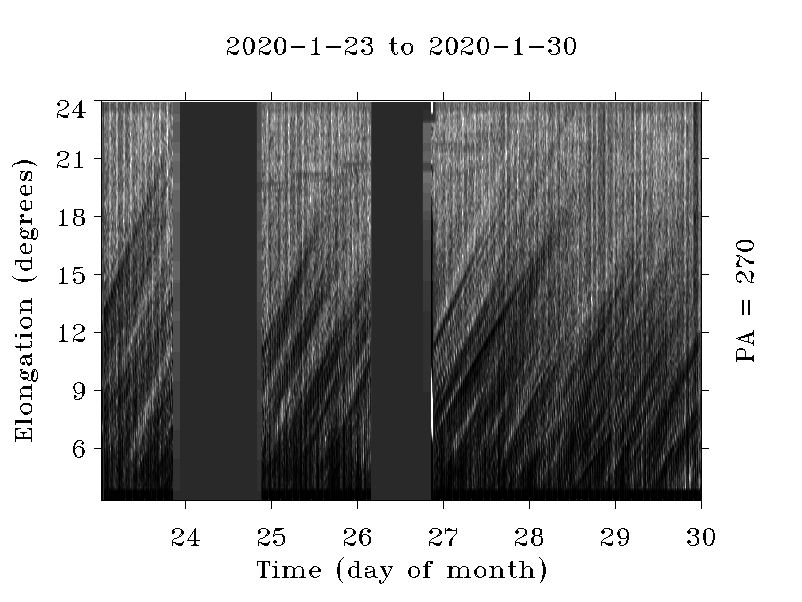}
\caption{STEREO-A HI1 J-map during PSP E4.}
\label{hi1}
\end{figure*}

We note here that  our methodology produces  
a composite J-map from data from both WISPR-I and WISPR-O. However, 
only the CME tracks can be traced through the full map.
This is likely
due to the lower signal-to-noise ratio of the fainter tracks in WISPR-O, since it is pointed further away from the Sun. 
Our ability to systematically track solar wind flows in WISRP-O J-maps should improve as solar activity increases and PSP perihelia move inward.

Finally, we discuss complementary observations from the SECCHI HI1
heliospheric imager  on-board the STEREO-A spacecraft. STEREO-A is following Earth at distance slightly smaller than 1 AU ($\approx$ 0.96 AU during E4; Fig. \ref{fov}). HI1 obtaines total brightness images of the corona and inner heliosphere in the 630-730 nm bandpass at a cadence of 40 minutes  with a FOV  from 4 to 24\dg\  elongation. During 24-26 January 2020
PSP and STEREO-A were radially aligned and the WISPR-I and HI1-A FOVs overlapped (see Fig. \ref{fov}, left panel). The two telescopes were viewing the
same region in the corona and inner heliosphere. In addition, the two spacecraft were in near-quadrature with Earth, so WISPR-I and
HI1-A traced plasma elements moving toward the Earth.

We constructed an HI1-A J-map for the E4 time interval by using
running differences of every two consecutive images and applying the
standard procedures described in \citet[][]{davies2009}. The
HI1 J-map corresponded to a rectangular 1\dg-wide slab along the Sun-Earth
line and is displayed in Fig. \ref{hi1}. Among the several tracks in this J-map, emphasis should be placed on the tracks during 25-26 January, which is the interval of  STEREO-A-PSP-Sun radial alignment of interest. 
During this interval, we counted similar number of tracks 
(about 4-5) in both HI1-A and WISRP-I J-maps.  We discuss these further in Section 5. 

\subsection{R-map}

The parabolic shape of the R-map (Fig. \ref{rmap}) represents the variation of the impact radius on the Thomson surface and hence the PSP radial distance. We can identify several tracks with positive slopes on the R-map. Linear fitting between the manually-selected start and end points yields speeds between 150 and 300  km s$^{-1}$. These speeds fall within the range of streamer blob speeds \citep[e.g.,][]{sheeley1999} and of slow CMEs  \citep[e.g.,][]{vourlidas2002}. Due to the rapidly changing viewpoint and heliocentric distance, the speed calculation for transient features in WISPR images is nontrivial \citep[see][]{liewer2019}. %The use of R-maps may provide an alternative path to streamline such calculations.

We chose the R-maps for the calculation of the outflow speeds because they represent the easiest and most conventional way to translate angular distances to linear distances in imaging observations of optically thin structures. The R-map method is based on the so-called ``Point P'' method \citep[e.g.,][]{Howard2006} that has been used for STEREO observations, by also taking into account the rapidly varying distances and viewpoints associated with the WISPR observations. Both methods make the underlying assumption that the structure is concentrated on the plane of the sky. There are several other methods to derive outflow kinematics \citep[e.g., see][and references therein]{Mishra2014} but in case of WISPR data they should be adjusted to incorporate the effect of spacecraft motion. When the appropriate modifications are in place, it will be straightforward to compare their results to R-map calculations but this is beyond the scope of this paper.

\subsection{Lat-map}

The Lat-map (Fig. \ref{latmap}) reveals a streamer bifurcation during the period of super-rotation (that is, from about January 27 to January 31; we note that the bright V-shaped feature corresponds to the CME) and close to
the perihelion on 29 January. A bright branch moves southward while the other dimmer branch remains close to the PSP orbit plane. This effect was first noted by \citet[][]{poirier2020} in a similar Lat-map during the first PSP encounter. The bifurcation and the difference in intensities indicate that PSP flew through a slightly folded streamer. The southern segment likely corresponds to either a denser sheet or a longer LOS (i.e., flatter sheet) or a sheet located nearest the Thomson surface or all of the above.  
The other branch could be fainter because PSP was approaching it during super-rotation or it was simply less dense or narrower. At this stage, it appears that the distance to PSP is the more likely explanation, as the branch seems to progressively brighten as the Sun overtakes PSP (that is, the spacecraft angular speed drops below the solar rotation speed and hence the Sun ``overtakes'' PSP in angular speed) at the end of the outbound corotation interval. More definitive interpretations  require more extensive analysis, which is beyond the scope of this paper. It should be clear, however, that 
detailed analyses of such maps will be helpful in reconstructing the 3D structure of streamers \citep[see the forward modeling of][]{poirier2020}.  Finally, during the two corotation periods (26-27 January and 1-2 February) the distance between WISPR and the streamer does not change appreciably, and therefore the Lat-map can be used to track the evolution of the internal streamer structure. During the first corotation period, the streamer is affected by the CME. There is much less variability during the second quasi-corotation period. 

\section{Conclusions-outlook}

We present a methodology for assessing and measuring the kinematics of propagating structures in the WISPR images. Our approach is based on well-established techniques developed for STEREO but it is tailored to address several technical challenges arising from the rapidly varying WISPR viewpoint. Our results are summarized as
follows:

Firstly, we introduce data products to assess the temporal evolution of coronal intensity as a function of elongation, projected distance (on the Thomson surface), and latitude. These J-map, R-map and Lat-map data products, respectively, comprise a novel framework to study transient solar wind flows observed from rapidly varying viewpoints such as PSP and Solar Orbiter.

Secondly, the reliability of the WISPR J-maps is validated against a movie of the WISPR images where a CME, a few blobs, and fronts could be identified, as well as against a ``raw'' J-map. 

Thirdly, several curved tracks of bright outflows (some of them marked in the bottom panel of Fig. 4) can be identified and traced in the J-map (4-5/day). They propagate at angular rates of $\approx$ 2.28 $\pm$ 0.7\dg/hour ($2.49 \pm 0.95$\dg/hour) for linear (quadratic) fittings.

Fourthly, the analysis of the R-map, indicates linear speeds in the range 150-300 \kms, which are consistent with past measurements of slow solar wind blobs and slow CMEs.

Fifthly, a comparison of the WISPR J-map against a STEREO-A HI1 J-map shows that a similar number of tracks can be identified in both J-maps during the interval of STA-PSP-Sun quasi-radial alignment.

Sixthly, the analysis of a Lat-map, which maps the variation of intensity as a function of latitude from the PSP orbit plane, shows evidence of ``flying-into'' a streamer during a PSP period of super-rotation. This representation of the data has many uses. The map can be used to detect faint CMEs and measure the width of brighter CMEs like the one passing though the FOV on 26 January 
\citep[details to be reported in an upcoming paper by][in this issue]{Liewer2020}. It can also be helpful in assessing the effects of CMEs on the background streamer structure and the evolution of the structure itself (e.g., bifurcation). 

Our methodology is able to recover, with significant  clarity, a plethora of tracks in the resulting J-map. The
daily number of $\sim$4-5 tracks is consistent with previous results on the statistics of streamer blobs during solar minimum conditions observed with LASCO and SECCHI \citep[e.g.,][]{wang1998,sheeley2009}. The WISPR detections are validated further by identifying a similar number of tracks in a co-temporal HI1 j-map (Fig. \ref{hi1}) during a period of overlapping FOVs between the two instruments.

We find, however, that the similar number of tracks  between WISPR and HI-1 is a bit unexpected. The WISPR lines-of-sight through coronal structures should be shorter than HI-1 lines-of-sight, given the proximity of PSP to the streamers and the smaller Thomson sphere for WISPR observations compared to that for SOHO and STEREO 1~AU observations. The shorter lines-of-sight should minimize structure overlap and should lead to the detection of more tracks, if they existed. On the other hand, the shorter lines-of-sight may contain fewer structures. Therefore, the similar rate of 4-5 blobs per day
from $< 55$ \rs\ and 1~AU is either a chance interplay between line-of-sight and overlap or a physical characteristic of the solar wind. Such a result may have important
implications for the global rate of generation of  transient solar wind flows and deserves further, more comprehensive studies, on which we will report soon.

In addition to the occurrence rates of streamer blobs, previous studies have highlighted the existence of periodic (electron) density structures (PDS) in SECCHI observations. Power spectrum analysis of PDS shows evidence of a wide range of periods, from 20 min to 19.5 hours \citep[][]{viall2010,telloni2013,viall2015,sanchez2017,deforest2018}. The longer periods are likely associated with larger-scale blobs while shorter periods could be associated with structures embedded within the larger streamer blobs. Indeed, a closer inspection of the J-map in Fig. \ref{jmap1} suggests that some of the tracks may not be monolithic but rather contain significant substructure. This is another study we wish to embark on.
Finally, future work will include the application of our methodology to construct J-maps and other pertinent maps to upcoming observations from SoloHI \citep[][]{Howard20} and METIS \citep[][]{Antonucci20} on board the Solar Orbiter 
\citep[][]{Muller20} mission. 

\begin{acknowledgements}
We thank the referee for his/her comments which led to improvement of the paper. AN and SP would like to thank Angelos Vourlidas for inviting them to the APL in summer 2019 where they were introduced to the analysis of PSP WISPR data. AV is supported by WISPR Phase-E funding to APL. We acknowledge the work of the PSP operations team. Parker Solar Probe was designed, built, and is now operated by the Johns Hopkins Applied Physics Laboratory as part of NASA’s Living with a Star (LWS) program (contract NNN06AA01C). We also acknowledge the work of the WISPR team developing and operating the instrument. The work of PCL, PP, and JRH was conducted at the Jet Propulsion Laboratory, California Institute of Technology under a contract from NASA.  
\end{acknowledgements}

\bibliographystyle{aa}
\bibliography{jmap_paper_accepted.bib}

\end{document}